\documentclass[english,aps,pra,reprint,superscriptaddress]{revtex4-1}
\usepackage{graphicx}
\usepackage{color}
\usepackage[colorlinks=true, pdfstartview=FitV, linkcolor=blue, citecolor=blue, urlcolor=blue]{hyperref}
\usepackage{multirow}

\begin{document}
\title{Interlayer magnetism in Fe3-xGeTe2}
\author{Xiangru Kong}
\affiliation{Center for Nanophase Materials Sciences, Oak Ridge National Laboratory, Oak Ridge, Tennessee 37831, USA}
\author{Giang D. Nguyen}
\affiliation{Center for Nanophase Materials Sciences, Oak Ridge National Laboratory, Oak Ridge, Tennessee 37831, USA}
\affiliation{Stewart Blusson Quantum Matter Institute, University of British Columbia, Vancouver, British Columbia V6T 1Z4, Canada}
\author{Jinhwan Lee}
\affiliation{School of Mechanical Engineering, Sungkyunkwan University, Suwon, Kyonggi-do 16419, Korea}
\author{Changgu Lee}
\affiliation{School of Mechanical Engineering, Sungkyunkwan University, Suwon, Kyonggi-do 16419, Korea}
\affiliation{SKKU Advanced Institute of Nanotechnology, Sungkyunkwan University, Suwon, Kyonggi-do 16419, Korea}
\author{Stuart Calder}
\affiliation{Neutron Scattering Division, Oak Ridge National Laboratory, Oak Ridge, Tennessee 37831, USA.}
\author{Andrew F. May}
\affiliation{Materials Science and Technology Division, Oak Ridge National Laboratory, Oak Ridge, TN 37831}
\author{Zheng Gai}
\affiliation{Center for Nanophase Materials Sciences, Oak Ridge National Laboratory, Oak Ridge, Tennessee 37831, USA}
\author{An-Ping Li}
\affiliation{Center for Nanophase Materials Sciences, Oak Ridge National Laboratory, Oak Ridge, Tennessee 37831, USA}
\author{Liangbo Liang}
\affiliation{Center for Nanophase Materials Sciences, Oak Ridge National Laboratory, Oak Ridge, Tennessee 37831, USA}
\author{Tom Berlijn}
\email{berlijnt@ornl.gov}
\affiliation{Center for Nanophase Materials Sciences, Oak Ridge National Laboratory, Oak Ridge, Tennessee 37831, USA}
\affiliation{Computational Sciences $\&$ Engineering Division, Oak Ridge National Laboratory, Oak Ridge, Tennessee 37831, USA}

\begin{abstract}
Fe$_{3-x}$GeTe$_2$ is a layered van der Waals magnetic material with a relatively high ordering temperature and large anisotropy. While most studies have concluded the interlayer ordering to be ferromagnetic, there have also been reports of interlayer antiferromagnetism in Fe$_{3-x}$GeTe$_2$. Here, we investigate the interlayer magnetic ordering by neutron diffraction experiments, scanning tunneling microscopy (STM) and spin-polarized STM measurements, density functional theory plus U calculations and STM simulations. We conclude that the layers of Fe$_{3-x}$GeTe$_2$ are coupled ferromagnetically and that in order to capture the magnetic and electronic properties of Fe$_{3-x}$GeTe$_2$ within density functional theory, Hubbard U corrections need to be taken into account. 
\end{abstract}
\maketitle

\section{introduction}

Magnetic layered materials, due to their reduced dimensionality, have exceptional physical properties and potential applications in spintronics~\cite{novoselov20162d,geim2013van,deng2018gate}.
Recently, Fe$_{3-x}$GeTe$_2$ has attracted much attention as a magnetic van der Waals layered material with a wide range of interesting properties~\cite{deiseroth2006fe3gete2,chen2013magnetic,verchenko2015ferromagnetic,leon2016magnetic,may2016magnetic,zhu2016electronic,zhuang2016strong,liu2017critical,yi2016competing,nguyen2018visualization,zhang2018emergence}.
Spin-polarized scanning tunneling microscope measurements have demonstrated the existence of skyrmionic bubbles in Fe$_{3-x}$GeTe$_2$~\cite{nguyen2018visualization}. This has been confirmed in a series of follow-up studies~\cite{ding2019observation,wang2019direct,park2019observation,wu2019n,you2019angular}.
In addition, several other intriguing physical properties have been discovered in Fe$_{3-x}$GeTe$_2$ , such as a large anomalous Hall current induced by topological nodal lines~\cite{kim2018large}, Kondo lattice behavior~\cite{zhang2018emergence}, giant tunneling magnetoresistance~\cite{song2018giant} and current-driven magnetization switching~\cite{wang2019current}.
It is noteworthy that the magnetic ordering temperature can be controlled by Fe occupancy~\cite{may2016magnetic} and gating~\cite{deng2018gate}. 
Furthermore, these materials have been cleaved to the monolayer limit where magnetic order persists and gating has been demonstrated to tune the Curie temperature order in exfoliated crystals~\cite{Fei2018, Deng2018}.

\begin{figure}[!htb]
\includegraphics[width=1.0\columnwidth]{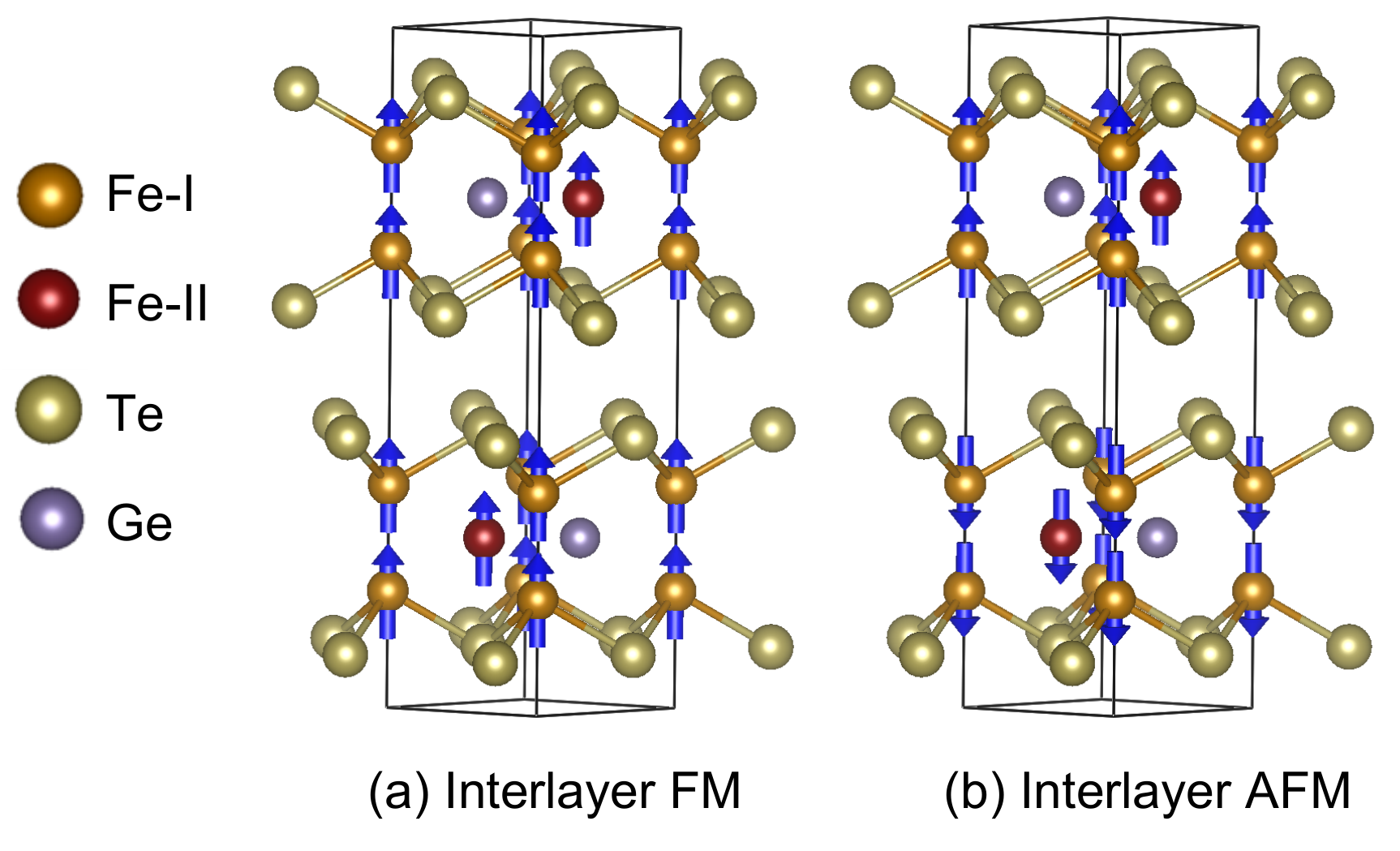}\caption{\label{fig:structure} Interlayer magnetic couplings between two layers in Fe$_{3-x}$GeTe$_2$. (a) FM (b) AFM.}
\end{figure}

As shown in Fig. $\text{\ref{fig:structure}}$, Fe$_{3}$GeTe$_2$ consists of a triangular layer of planar FeGe (with iron atoms
labeled Fe-II) sandwiched between two triangular layers of buckling FeTe (with iron atoms labeled Fe-I). The Fe$_{3}$GeTe$_2$ ``sandwiches'' are weakly bound to each other via van der Waals interactions. Interestingly it has been found that there can be large amounts of Fe-II vacancies in the FeGe layer up to 30\% ~\cite{verchenko2015ferromagnetic,may2016magnetic}. 

Bulk Fe$_{3-x}$GeTe$_2$ has been reported to be a van der Waals magnetic material with a relatively high magnetic ordering temperature and large easy-axis anisotropy~\cite{deiseroth2006fe3gete2,chen2013magnetic}.  Importantly, however, both the magnetic ordering temperature ($140 \;{\rm K} < T_c < 230 \;{\rm K}$) and magnetic anisotropy are reduced with increasing concentration of Fe-II vacancies~\cite{may2016magnetic}. Early on it was concluded that Fe$_{3-x}$GeTe$_2$ has an interlayer ferromagnetic (FM) ordering as shown in Fig. $\text{\ref{fig:structure}}$(a) based on magnetization measurements~\cite{deiseroth2006fe3gete2,chen2013magnetic,verchenko2015ferromagnetic,leon2016magnetic,may2016magnetic,zhu2016electronic,zhuang2016strong,liu2017critical}. Also the temperature dependence of several neutron diffraction peaks are consistent with FM ordering with the moments aligned along the $c$-axis~\cite{may2016magnetic}. However, it was found that when using a lower magnetic field, the magnetic susceptibility of Fe$_{3-x}$GeTe$_2$ suggest antiferromagnetic (AFM) ordering between the layers ( Fig. $\text{\ref{fig:structure}}$ (b)) instead of the FM ordering.~\cite{yi2016competing}. In particular, the temperature dependence of the magnetization shows a kink around a characteristic temperature $T^{*}$ below the ordering temperature $T_c$, with $T^{*}\sim150$ K and $T_c\sim210$ K for the samples used in Ref.~\cite{yi2016competing}. In addition the zero-field cooled (ZFC) magnetization curve goes very close to zero as the temperature goes to zero when using an external magnetic field parallel to the easy axis. Furthermore it was concluded from AC susceptibility measurements and magnetic force microscopy that in the temperature range between $T^{*}$ and $T_c$ the system enters a second more complex magnetic phase~\cite{yi2016competing}. To further support this scenario it was shown from Density Functional Theory (DFT) that at zero temperature the AFM interlayer ordering has a lower energy than the FM one~\cite{yi2016competing}.
A recent theoretical study ~\cite{jang2019origin} concluded that while undoped Fe$_{3}$GeTe$_2$ has an AFM interlayer ordering, the ground state of hole-doped Fe$_{3-x}$GeTe$_2$ ($0.11<x<0.36$) should have an interlayer FM ordering. The interlayer magnetic ordering controls inversion symmetry and time-reversal symmetry, which in turn controls many physical properties
such as the topology, giant magnetoresistance, second harmonic generation and piezo electricity~\cite{Li2019MnBi2Te4,klein2018probing,Sun2019CrI3SHG}.
Therefore it is important to resolve the interlayer magnetic ordering in Fe$_{3-x}$GeTe$_2$.
Understanding the interlayer coupling in Fe$_{3-x}$GeTe$_2$ may also help researchers to understand the magnetism in related metallic ferromagnets, notably Fe$_{5-x}$Ge$_2$Te$_2$~\cite{ Jothi2019} and Fe$_{5-x}$GeTe$_2$~\cite{Stahl2018,May2019_ACSNano, May2019_PRM}, the latter of which has a Curie temperature over 300K.

In our work, a series of experimental and theoretical methods has been employed to study the interlayer magnetic ordering of Fe$_{3-x}$GeTe$_2$. This includes neutron diffraction, scanning tunneling microscopy (STM) around Fe-II vacancies, spin-polarized STM measurements across step-edges and Density Functional Theory (DFT) calculations. Our results indicate that Fe$_{3-x}$GeTe$_2$ is ferromagnetically ordered between the layers. In order to reach agreement of the theoretical results with the experimental observations we have found that the DFT+U method is needed to 
treat the local Coulomb interactions among the Fe-d orbitals. 

\section{Methods}

\textit{Neutron diffraction experiments and samples}
Neutron scattering was performed on a 5 gram powder sample of composition Fe$_{2.9}$GeTe$_2$ and a single crystal sample of concentration Fe$_{2.76}$GeTe$_2$ with dimensions up to 10 mm. The preparation and characterization of these samples are described in Ref.~\cite{may2016magnetic}. The powder sample was loaded into a cylindrical Al can and measured on the HB-2A neutron diffractometer at the High Flux Isotope Reactor (HFIR), ORNL ~\cite{calder2018}. A constant wavelength of 2.41 \AA{} was selected from the 113 reflection of a vertically focusing Ge monochromator. The pre-mono, pre-sample and pre-detector collimation was open-21’-12’. Measurements were performed in the range 5 to 130 degrees. Single crystal elastic measurements were performed on the HB-3 triple axis instrument at HFIR using a PG(002) monochromator and analyzer with a fixed energy of 14.7 meV. The collimation was 48’-40’-40’-120’ for pre-mono, pre-sample, pre-analyzer and pre-detector. We note that the neutron beam is larger than the samples. We also note that the samples used for neutron scattering in this study correspond to some of the samples used in the previous literature. Specifically, the powder sample was measured in Ref. ~\cite{may2016magnetic} and the single crystal was the one discussed in Ref. ~\cite{yi2016competing}.

\textit{Scanning tunneling microscope and samples}
STM and spin-polarized STM was performed using a variable-temperature Omicron STM instrument. The sample was grown using the chemical vapor transport method as described in the Ref. ~\cite{nguyen2018visualization}. A single crystal of Fe$_{3-x}$GeTe$_2$ was cleaved \textit{in situ} in an ultrahigh vacuum at room temperature and then transferred to STM for measuring at 120 K. The atomic-resolution topographic image was acquired using a tungsten tip. Etched ferromagnetic nickel tips were prepared for spin-polarized STM measurements as in our previous reports~\cite{nguyen2018visualization,park2017}. d$I$/d$V$ maps were obtained using a lock-in technique with ac modulation of 1 kHz and 30 mV.

\textit{Theoretical methods.} DFT~\cite{hohenberg1964inhomogeneous} within the projected augmented wave method~\cite{kresse1996efficient,kresse1999ultrasoft} has been employed, as implemented in the Vienna \textit{ab initio} simulation package. The generalized gradient approximation of Perdew, Burke and Ernzerhof was used for the electron exchange-correlation functional~\cite{perdew1996generalized}. To account for the strong intra-atomic interactions two types of DFT+U approximation schemes were used ~\cite{dudarev1998electron, liechtenstein1995prb}. Also approximate van der Waals corrections have been added to treat the interactions between the layered structures, according to the optB86b-vdW method~\cite{klimevs2009chemical,klimevs2011van}. The in-plane and out-of-plane lattice constants $a=3.991$ \AA{} and $c=16.336$ \AA{}, the space group and the initial atomic positions were taken from X-ray diffraction~\cite{deiseroth2006fe3gete2}. The relaxation of the atom positions was performed by the conjugate gradient scheme until the maximum force on each atom was less than 1 meV\AA$^{-1}$, and the total energy was converged to $10^{-7}$ eV. The energy cutoff of the plane waves was chosen as 520 eV. To simulate the effects of Fe-II vacancies three type of supercells have been considered in which the in-plane lattice vectors have been extended, while keeping the out-of-plane lattice vectors fixed. These supercells are depicted in the supplement~\cite{supp}. The k-point grid used for the normal cell, the $\sqrt{3}\times\sqrt{3}$, $2\times2$ and $4\times1$ supercells were $12\times12\times3$, $7\times7\times3$, $5\times5\times3$ and $3\times12\times3$, respectively. To perform the STM simulations a vacuum of $\sim20$ \AA{} was inserted. To simplify the STM simulations, the atomic positions were taken to be those corresponding to the bulk supercells. The STM simulations have been performed with the P4VASP software~\cite{p4vasp} using the constant current mode. Atomic images were produced with the VESTA program~\cite{vesta}.

\section{Results}

\begin{figure}[!htb]
\includegraphics[width=1.0\columnwidth]{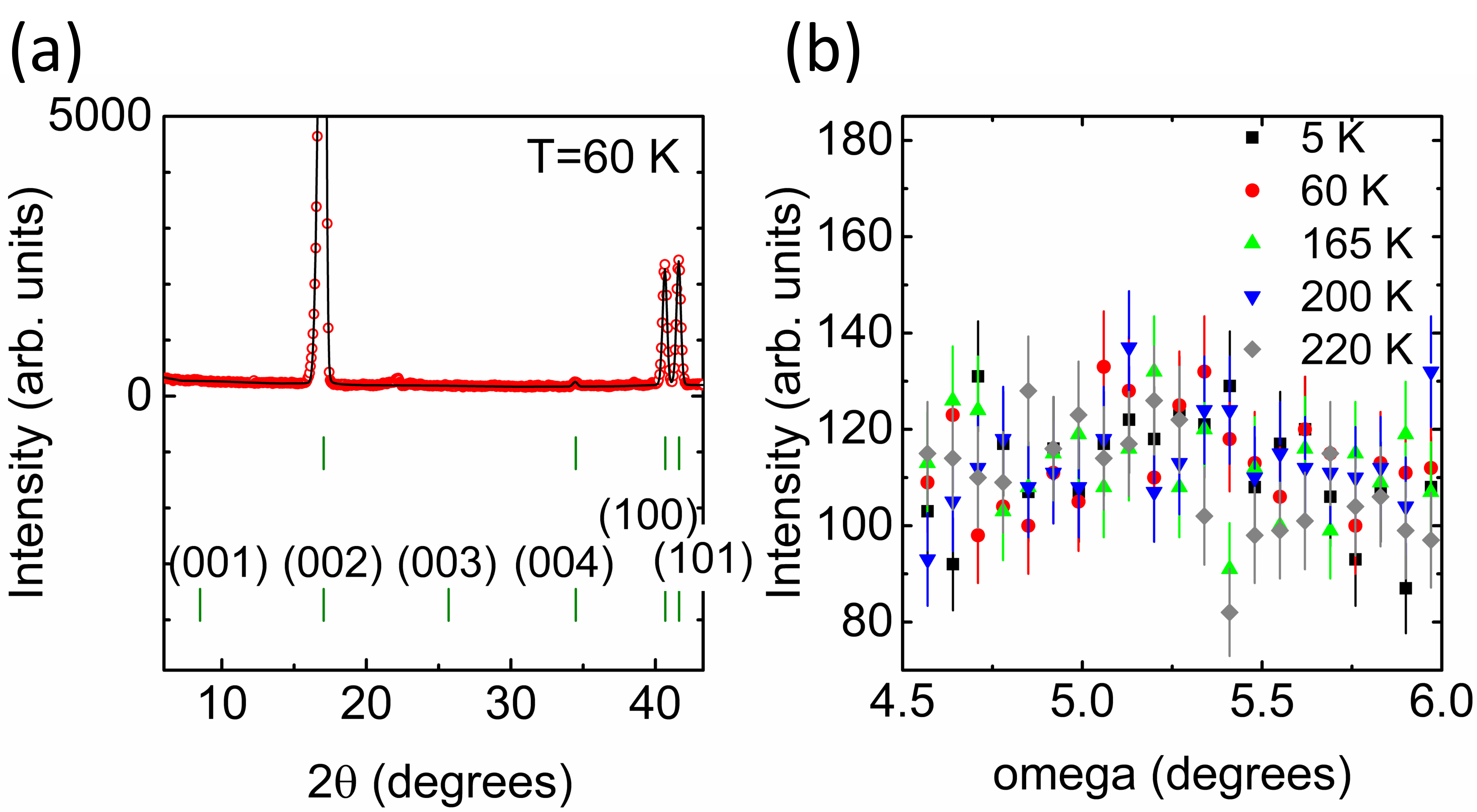}\caption{\label{fig:neutron}(a) Neutron powder diffraction at 60 K on Fe$_{2.9}$GeTe$_2$. (b) Single crystal measurements at the (003) reflection for temperatures of 5 K, 60 K, 165 K, 200 K and 220 K on Fe$_{2.76}$GeTe$_2$.}
\end{figure}

First, to investigate the possibility of interlayer AFM, neutron diffraction experiments were performed as shown in Fig. \ref{fig:neutron}. As we can see from the neutron diffraction in Fig. \ref{fig:neutron}(a) there are no Bragg peaks associated with AFM ordering at (001) and (003). These measurements were performed at 60 K which is well below $T^{*}\sim150$ K for which the kink was observed in the magnetization measurements of Fe$_{3-x}$GeTe$_2$ ~\cite{yi2016competing}. In addition the (003) refection in Fig. \ref{fig:neutron}(b) shows no significant temperature dependence in going from above the magnetic transition at 220 K down to 5 K. These results show there is no interlayer AFM long range order in Fe$_{3-x}$GeTe$_2$. 

\begin{figure}[!htb]
\includegraphics[width=1.0\columnwidth]{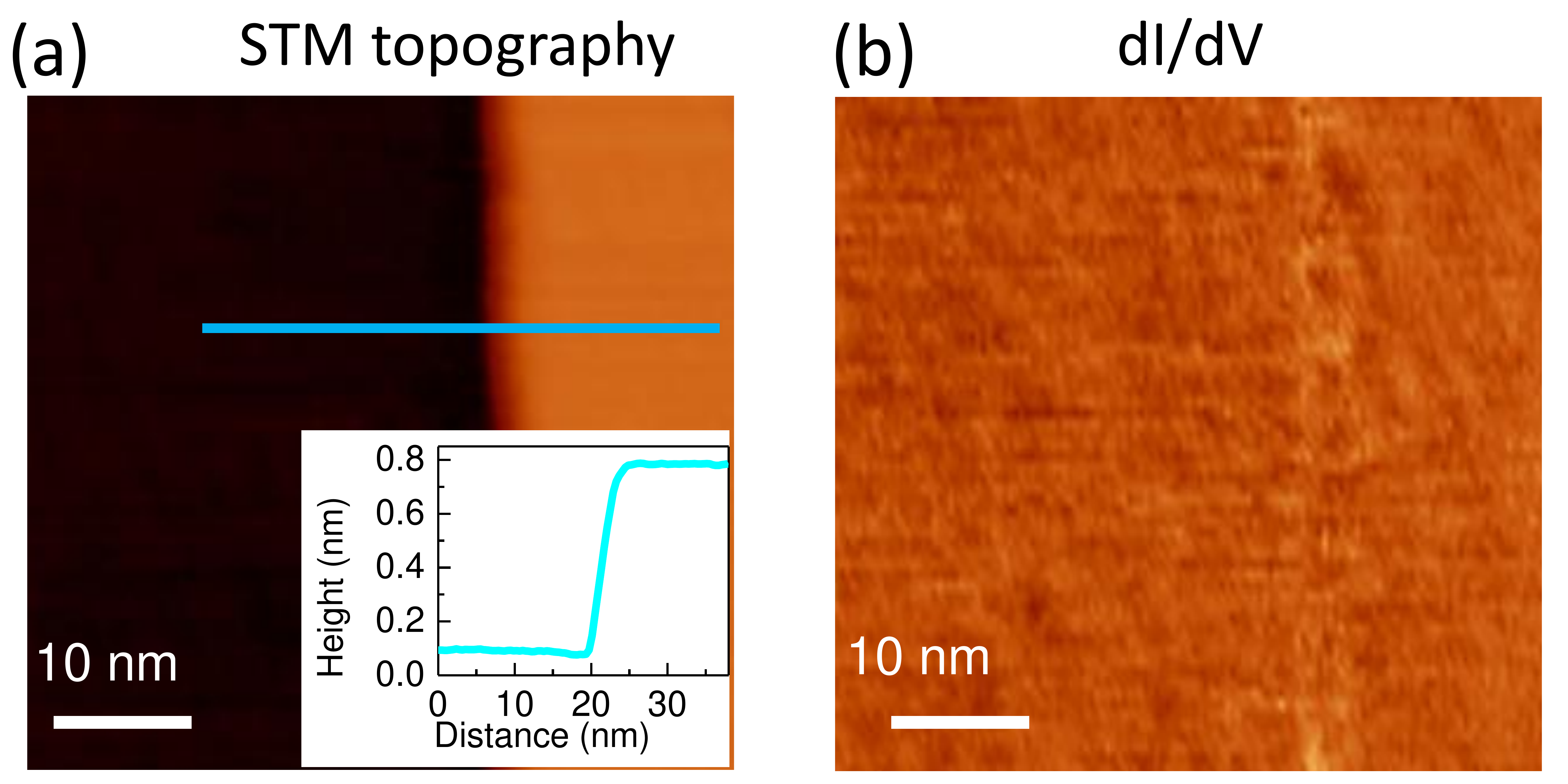}\caption{\label{fig:sp-stm} 
(a) STM topographic image of a stepped area with a line profile across the step (inset) (sample bias $V_{s}=-0.1$ V, tunneling current $I= 50$ pA) (b) The corresponding spin-polarized differential conductance (d$I$/d$V$) map which shows no contrast between two terraces.}
\end{figure}

In order to further study the possibility of interlayer AFM order we performed spin-polarized STM measurements across a large number of step-edges. A typical example is shown in Fig $\text{\ref{fig:sp-stm}}$. The height of a stepped layer is measured to be about 0.7 nm as shown in Fig $\text{\ref{fig:sp-stm}}$(a), which is roughly consistent with the half of lattice constant in $c$-axis ($\sim 0.8$ nm). Importantly, the map of d$I$/d$V$ at bias -0.1 V as shown in Fig $\text{\ref{fig:sp-stm}}$(b) shows no spin-polarized contrast between the two terraces. This is also confirmed by the d$I$/d$V$ maps at biases -0.5 V, -0.2 V and 0.2 V shown in the supplement~\cite{supp}. On the other hand, using the same STM tip, we were able to resolve magnetic domains in Fe$_{3-x}$GeTe$_2$ ~\cite{supp}. We note that no contrast in spin-polarized STM does not necessarily imply the same magnetic configuration. It could also be that the spin-polarized STM is not sensitive enough to detect a difference.  Nevertheless, the spin-polarized STM observation is consistent with the neutron scattering measurements discussed above, which show no evidence for AFM coupling between the layers. 

\begin{table}[!htb]
\begin{tabular}{|c|c|c|c|c|}
\hline
\multirow{2}{*}{U(eV)} & \multicolumn{4}{c|}{ Energy difference $E_{FM}-E_{AFM}$ (meV)} \\ 
\cline{2-5}
 & $1\times1$ cell & $\sqrt{3}\times\sqrt{3}$ cell & $2\times2$ cell & $4\times 1$ cell \\
\hline
0 & 22.4 & 16.0 & -5.0 & 35.0 \\
\hline
1 & -31.6 & -47.3 & -97.6 & -84.2 \\
\hline
2 & -45.9 & -51.1 & -112.4 & -116.8 \\
\hline
3 & -59.4 & -85.6 & -101.7 & -72.4 \\
\hline
4 & -41.1 & -91.1 & -188.1 & -176.8 \\
\hline
\end{tabular}\caption{\label{tab:energy}The total energy difference per cell between interlayer FM and AFM ordering in the normal cell without Fe-II vacancies and the single Fe-II vacancy $\sqrt{3}\times\sqrt{3}$, $2\times2$ and $4\times1$ supercells defined in the supplement~\cite{supp}, with the Hubbard U parameter increasing from 0 to 4 eV.}
\end{table}

Following the experimental results, we performed first-principle calculations to investigate the interlayer magnetic coupling in Fe$_{3-x}$GeTe$_2$. Several reports~\cite{zhang2018emergence,zhu2016electronic,xu2020spectroscopic} have suggested that correlations in Fe$_{3-x}$GeTe$_2$ play an important role. Given that Fe is a 3d transition metal this is a reasonable assertion. The simplest approach to treat local Coulomb interactions beyond DFT is via the so-called DFT+U method in which the Hubbard U parameter quantifies the local Coulomb repulsion among the Fe-d orbitals. To this end we first consider the DFT+U scheme implemented by Dudarev {\it et al} ~\cite{dudarev1998electron}. Tab. ~\ref{tab:energy} shows the energy difference between the interlayer FM and AFM configurations as a function of U. The second column corresponds to the Fe$_3$GeTe$_2$ normal cell without Fe-II vacancies. When $U=0$ eV, the energy difference between FM and AFM configuration for the normal cell without vacancies is $22.4$ meV, which is consistent with the DFT without U calculation reported in Ref. ~\cite{yi2016competing}. However, as we increase U from 0 to 4 eV, the energy difference $E_{FM}-E_{AFM}$ gives negative values as shown in the second column of Table $\text{\ref{tab:energy}}$. The negative energy difference for the listed finite U values show that the ground state of Fe$_3$GeTe$_2$ has an interlayer FM ordering for the case without vacancies, consistent with the experimental observations. When using the DFT+U scheme implemented by Liechtenstein {\it et al} ~\cite{liechtenstein1995prb}, the same conclusion is reached.~\cite{supp}

In Ref. ~\cite{jang2019origin} it has been concluded that Fe-II vacancies are responsible for inducing the ferromagnetism in Fe$_{3-x}$GeTe$_2$. To study the effect of Fe-II vacancies we considered the $\sqrt{3}\times\sqrt{3}$, $2\times2$ and $4\times1$ single Fe-II vacancy supercells defined in the supplement~\cite{supp}. Tab. ~\ref{tab:energy} shows the total energy of the FM configuration per supercell relative to that of the AFM configuration for these various supercells. Let us first focus on the $U=0$ results. Just like in Ref. ~\cite{jang2019origin} we find that for the $2\times2$ supercell the Fe-II vacancy flips the  FM configuration from being energetically unfavorable to favorable. However, for the $\sqrt{3}\times\sqrt{3}$ and $4\times 1$ supercells we find that the Fe-II vacancies don't reduce the total energy of the FM configuration relative to that of the AFM configuration enough to stabalize it. Therefore it seems we cannot draw a simple conclusion about the influence of Fe-II vacancies on the stability of the FM and AFM configurations. On the other hand, we see from Tab. ~\ref{tab:energy} that for all the listed finite values of U the FM configuration has the lowest energy in all the three supercells, in agreement with the experimental observations. 

\begin{figure}[!htb]
\includegraphics[width=1.0\columnwidth]{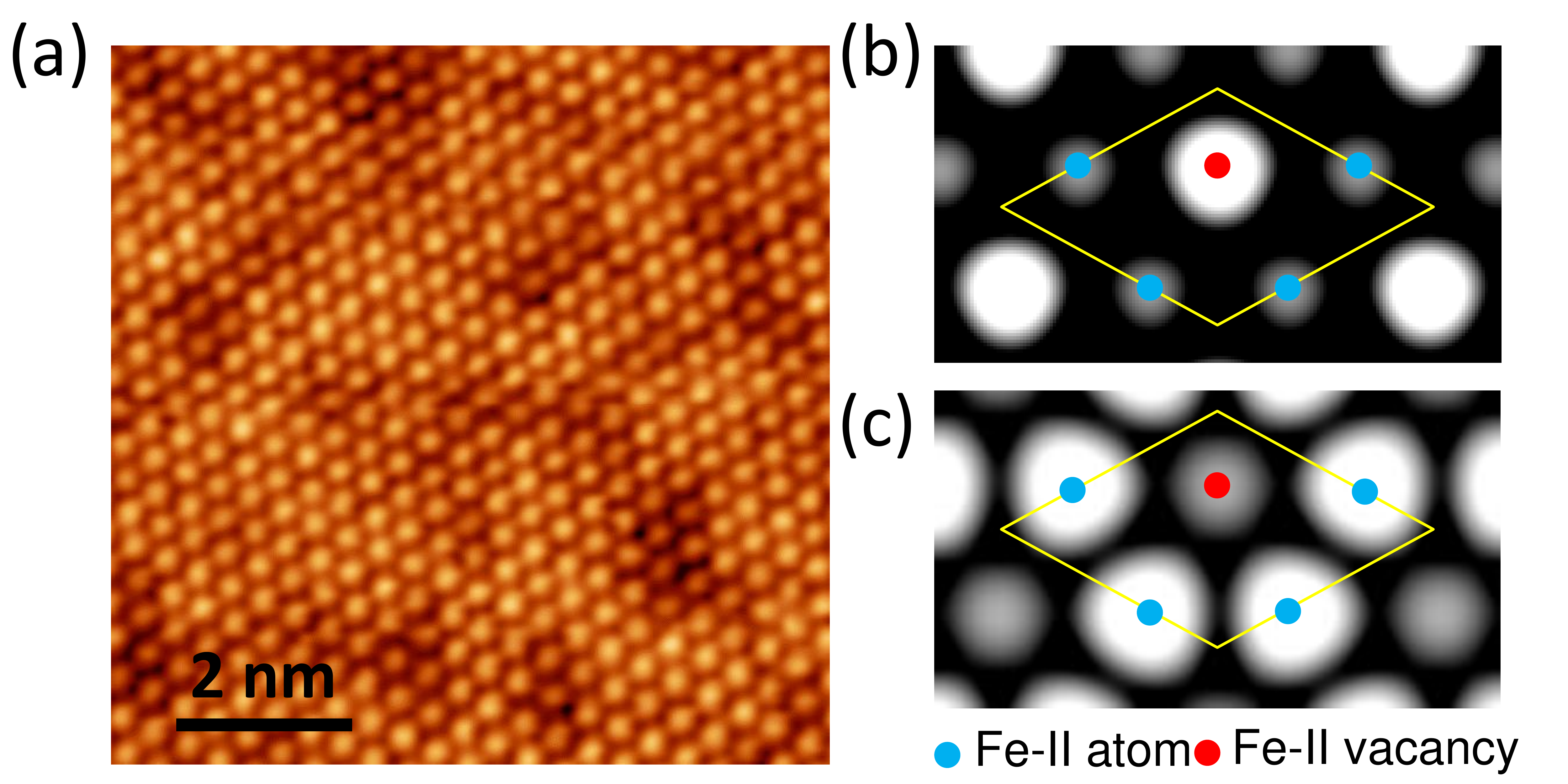}\caption{\label{fig:stm}(a) Experimental STM image of Fe$_{3-x}$GeTe$_2$ (tunneling current $I= 100$ pA) and theoretical DFT+U  STM simulation of a $\sqrt{3}\times\sqrt{3}$ single Fe-II vacancy supercell for (b) the AFM solution with U=0 and (c) the FM solution with $U=4$ eV. In both experiment and simulation the bias $V_{s}=-0.5$ V is used.}
\end{figure}

Finally we find that the DFT+U approximation is also more consistent with recent STM  experiments compared to plain DFT. Naturally, Fe$_{3-x}$GeTe$_2$ is Te terminated, so the STM tip is probing the local electronic structure above the Te atoms. The Te atoms in turn are located directly above the Fe-II atoms. In Ref. ~\cite{nguyen2018visualization} it had been shown that the STM intensities above the Fe-II vacancies is darker than those above the Fe-II atoms. In Fig.$\text{\ref{fig:stm}}$(a) a similar experimental STM image is presented that shows the same effect. To simulate the STM around Fe-II vacancies we perform DFT+U calculations on a $\sqrt{3}\times\sqrt{3}$ supercell with a single Fe vacancy in it. Fig.$\text{\ref{fig:stm}}$(c) shows that the contrast  seen in STM is reproduced by the FM solution for $U=4$eV, because the STM intensity above the Fe-II vacancy is lower compared to the sites that are above the Fe-II atoms. On the other hand the STM simulation based on the AFM solution for $U=0$, shown in Fig.$\text{\ref{fig:stm}}$(b), has the opposite contrast: the STM intensity is bright/dark above the Fe-II vacancies/atoms. In other words, the STM simulation based on DFT without U corrections qualitatively disagrees with the STM experiments. In the supplement we demonstrate the same conclusion for the $2\times2$ and $4\times1$ single Fe-II vacancy supercells~\cite{supp}. 

To better understand the origin of the STM contrast above the Fe-II vacancies and Fe-II atoms we analyzed the relaxed positions and local density of states of the Te atoms that are located directly above the Fe-II atoms. For all the considered Fe-II vacancy supercells and values of the Hubbard U, in both the FM and AFM configuration, the Te heights above the Fe-II vacancies are lower than the Te heights above the Fe-II atoms~\cite{supp}.  So even for the AFM DFT case without a Hubbard U correction we find that the Te height above the Fe-II vacancy is lowered despite the fact that the STM intensity above these Te atoms is raised. This suggest that the origin of the contrast in the STM is electronic in nature instead of structural. To study the electronic contributions to the STM intensity we consider the local density of states (DOS) of the Te-$p$ orbitals integrated between the bias voltage of -0.5 eV used in the STM experiment and the Fermi energy. For all three supercells we find that for the AFM DFT simulation without a Hubbard U correction the integrated DOS of the Te-$p$ orbitals above the Fe-II vacancies is higher than the integrated DOS of the Te-$p$ orbitals above the Fe-II atoms. For the FM DFT+U simulations with $U=4$ eV this is reversed ~\cite{supp}. This further confirms that indeed the contrast in the STM simulations is driven by changes in the local electronic structure. 

\section{Discussion}

In our work we find that the interlayer magnetic coupling  in Fe$_{3-x}$GeTe$_2$ is FM and that in order to capture this theoretically, Hubbard U corrections need to be included within DFT. This conclusion is consistent with some earlier studies that highlight the importance of correlation effects in Fe$_{3-x}$GeTe$_2$ ~\cite{zhang2018emergence,zhu2016electronic,xu2020spectroscopic}. While we have focused on the interlayer magnetic coupling in our study, our conclusion that local Coulomb interactions play an important role in Fe$_{3-x}$GeTe$_2$ will help understanding the properties of this system in general. 

One question that remains unanswered is why the DFT calculations show that upon including a Hubbard U correction, Fe$_{3-x}$GeTe$_2$ undergoes an AFM to FM transition. More generally this ties to the question what the nature is of the FM interlayer coupling. In the van der Waals system CrI$_3$ it has been concluded that the interlayer magnetic coupling is controlled by the super-superexchange mechanism ~\cite{nikil2018CrI3}. However, unlike CrI$_3$, Fe$_{3-x}$GeTe$_2$ is metallic. In particular, the in-plane as well as the out-of-plane resistivity does not show activation behavior as a function of temperature. ~\cite{kim2018large}. At the same time there have been several reports with evidence for the presence of local moments in Fe$_{3-x}$GeTe$_2$ ~\cite{zhang2018emergence,zhu2016electronic,xu2020spectroscopic, calder2019prb}. One possibility is that within pure DFT the AFM is driven by nesting conditions in the out-of-plane band structure and that within DFT+U a FM exchange mechanism based on local moments becomes dominant. Since the AFM coupling occurs between layers within the unit cell, the nesting properties cannot directly be analyzed from the band structure. Instead one can compute the Lindhard function to study a spin-wave density scenario as was done for example for the case of intra-unitcell AFM in RuO$_2$~\cite{berlijn2017RuO2}. We leave such an endeavor for future studies.  

Another question is how the magnetization measurements can be interpreted given that there is no AFM in Fe$_{3-x}$GeTe$_2$. Specifically in Ref.~\cite{yi2016competing} it has been shown that the ZFC magnetization curve for low magnetic fields along the easy axis is very close to zero at low temperatures. This is indicative of AFM~\cite{Kittel}.  Furthermore, the magnetization curves display a kink around a characteristic temperature $T^{*}$ that is below the magnetic ordering temperature $T_c$. In Ref. ~\cite{cktian2019domainwall} these findings in the magnetization measurements have been reproduced. However, the authors of Ref. ~\cite{cktian2019domainwall} argue that domain-wall pinning is a more likely explanation for the observed behavior than AFM. After cooling Fe$_{3-x}$GeTe$_2$ in zero external field it contains an equal number of domains parallel and anti-parallel to the easy axis. Then if the magnetic fields and the temperatures are too low, the walls between these domains remain pinned and the induced magnetization stays negligible. The kink at $T^{*}$ is interpreted in Ref. ~\cite{cktian2019domainwall} as the temperature at which the domain walls depin. Magnetic force microscopy measurements have shown that the magnetic domains evolve from a branching structure at $T_c$ to a bubble structure at $T^{*}$ and remain constant at lower temperatures~\cite{yi2016competing}. It would be interesting to see if in future studies these observations can be understood within a domain-wall depinning scenario.

\section{Conclusions}

Using neutron diffraction, scanning tunneling microscopy, spin-polarized STM and DFT+U based total energy calculations and STM simulations we have studied the question whether the magnetic interlayer ordering in Fe$_{3-x}$GeTe$_2$ is FM or AFM. Based on our experimental results and the previous results in the literature we conclude that Fe$_{3-x}$GeTe$_2$ is FM. The resolution of the out-of-plane magnetic order is important given that it controls the time-reversal and inversion symmetry and therefore many physical properties such as the topology, second harmonic generation and piezo electricity to mention but a few. To theoretically reproduce the experimental observations we have learned that it is necessary to include Hubbard U corrections beyond pure DFT. This conclusion derived from the interlayer magnetic coupling will help understanding the electronic and magnetic properties of Fe$_{3-x}$GeTe$_2$ in general.

\begin{acknowledgments}
We thank R. Fishman for valuable discussions. This research was conducted at the Center for Nanophase Materials Sciences, which is a DOE Office of Science User Facility. We used resources of the Compute and Data Environment for Science (CADES) at the Oak Ridge National Laboratory, which is supported by the Office of Science of the U.S. Department of Energy under Contract No. DE-AC05-00OR22725. In addition we used resources of the National Energy Research Scientific Computing Center, a DOE Office of Science User Facility supported by the Office of Science of the U.S. DOE under Contract No. DE-AC02-05CH11231. This research used resources at the High Flux Isotope Reactor, a DOE Office of Science User Facility operated by the Oak Ridge National Laboratory. Sample synthesis (A.F.M.) was supported by the U. S. Department of Energy, Office of Science, Basic Energy Sciences, Materials Sciences and Engineering Division.

This manuscript has been authored by UT-Battelle, LLC under Contract No. DE-AC05-00OR22725 with the U.S. Department of Energy. The United States Government retains and the publisher, by accepting the article for publication, acknowledges that the United States Government retains a non-exclusive, paid-up, irrevocable, world-wide license to publish or reproduce the published form of this manuscript, or allow others to do so, for United States Government purposes. The Department of Energy will provide public access to these results of federally sponsored research in accordance with the DOE Public Access Plan (http://energy.gov/downloads/doe-public-access-plan).

\end{acknowledgments}

\begin{widetext}
\clearpage
{\Large\bf [Supplementary Information] Interlayer magnetism in Fe3-xGeTe2}

\setcounter{equation}{0}
\setcounter{figure}{0}
\setcounter{table}{0}
\setcounter{page}{1}
\setcounter{section}{0}
\makeatletter
\renewcommand{\theequation}{S\arabic{equation}}
\renewcommand{\thefigure}{S\arabic{figure}}
\renewcommand{\bibnumfmt}[1]{[S#1]}
\renewcommand{\citenumfont}[1]{S#1}
\renewcommand{\thetable}{S\Roman{table}}

\section{Spin-polarized scanning tunneling spectroscopy}\label{app:otherbias}

Fig. \ref{fig:otherbias}(a)-(c) show additional spin-polarized differential conductance (d$I$/d$V$) maps across the step edge at various biases. Similar to the one in Fig. 3(b) in the manuscript, these spin-polarized maps do not display a magnetic contrast between the terraces. The bright-line feature at the step edge is likely corresponding to edge states. Fig. \ref{fig:otherbias}(d) shows a spin-polarized STM topography image that reveals magnetic domains on the flat terrace, similar as seen in Ref.~\cite{nguyen2018visualization}.

\begin{figure}[!htb]
\includegraphics[width=1.0\columnwidth]{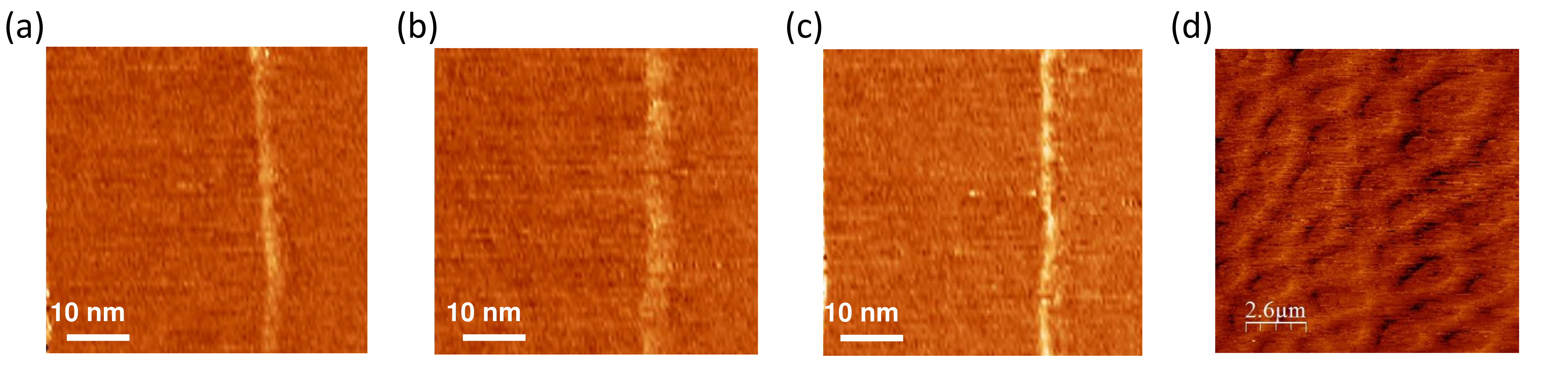}\caption{\label{fig:otherbias}
(a)(b)(c) Spin-polarized differential conductance (dI/dV) maps of the step-edge at biases $V_s$ -0.5, -0.2 and 0.2 V and tunneling current $I= 50$ pA. (d) Spin-polarized STM topography image of magnetic domains (sample bias $V_{s}=-0.5$ V, tunneling current $I= 50$ pA).}
\end{figure}

\clearpage
\section{Fe-II vacancy supercells}

Fig. ~\ref{fig:supercells} shows the supercells used in this study to simulate Fe-II vacancies. 

\begin{figure}[!htb]
\includegraphics[width=0.8\columnwidth]{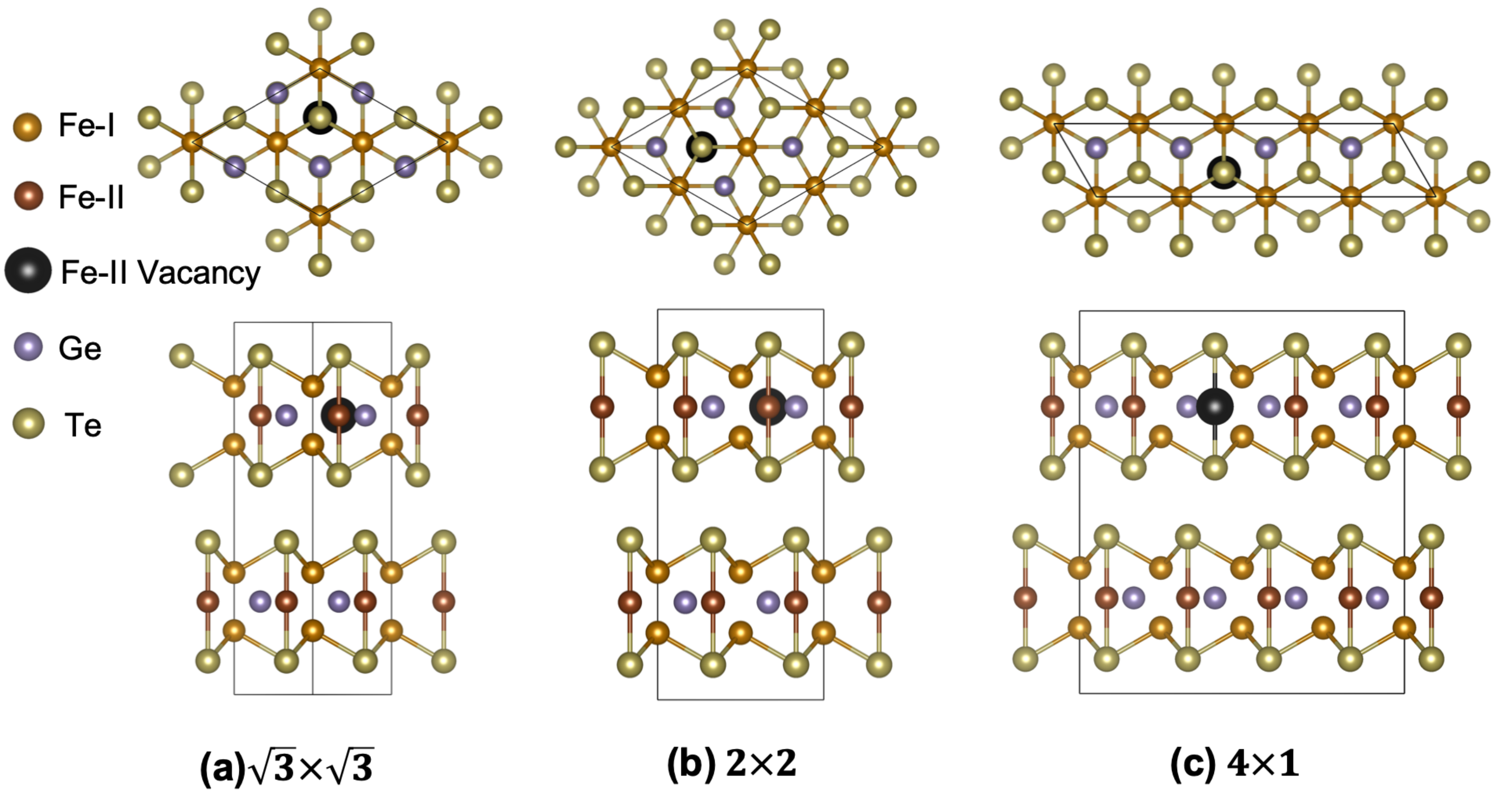}\caption{\label{fig:supercells}
Top and side view of three supercells used to simulate Fe-II vacancies.}
\end{figure}

\clearpage
\section{STM simulations}

Fig. ~\ref{fig:otherstm1} displays STM simulations for the $\sqrt{3}\times\sqrt{3}$ Fe-II vacancy supercell in the magnetic groundstate configuration as a function of U. These images show how the STM intensity above the Fe-II vacancy, relative to the STM intensity above the Fe-II atoms, changes from bright to dark upon increasing U. Fig. ~\ref{fig:otherstm2} shows the same for two other supercells.

\begin{figure}[!htb]
\includegraphics[width=1.0\columnwidth]{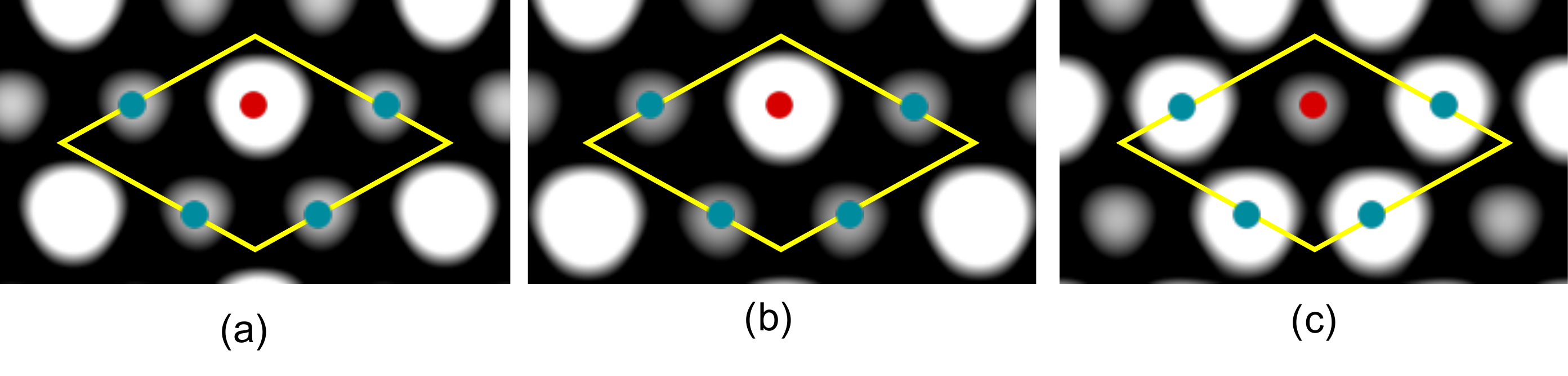}\caption{\label{fig:otherstm1}
STM simulations for the $\sqrt{3}\times\sqrt{3}$ Fe-II vacancy supercell in the magnetic groundstate configuration as a function of U: (a)U=1 eV; (b)U=2 eV; (c)U=3 eV.}
\end{figure}

\begin{figure}[!htb]
\includegraphics[width=0.98\columnwidth]{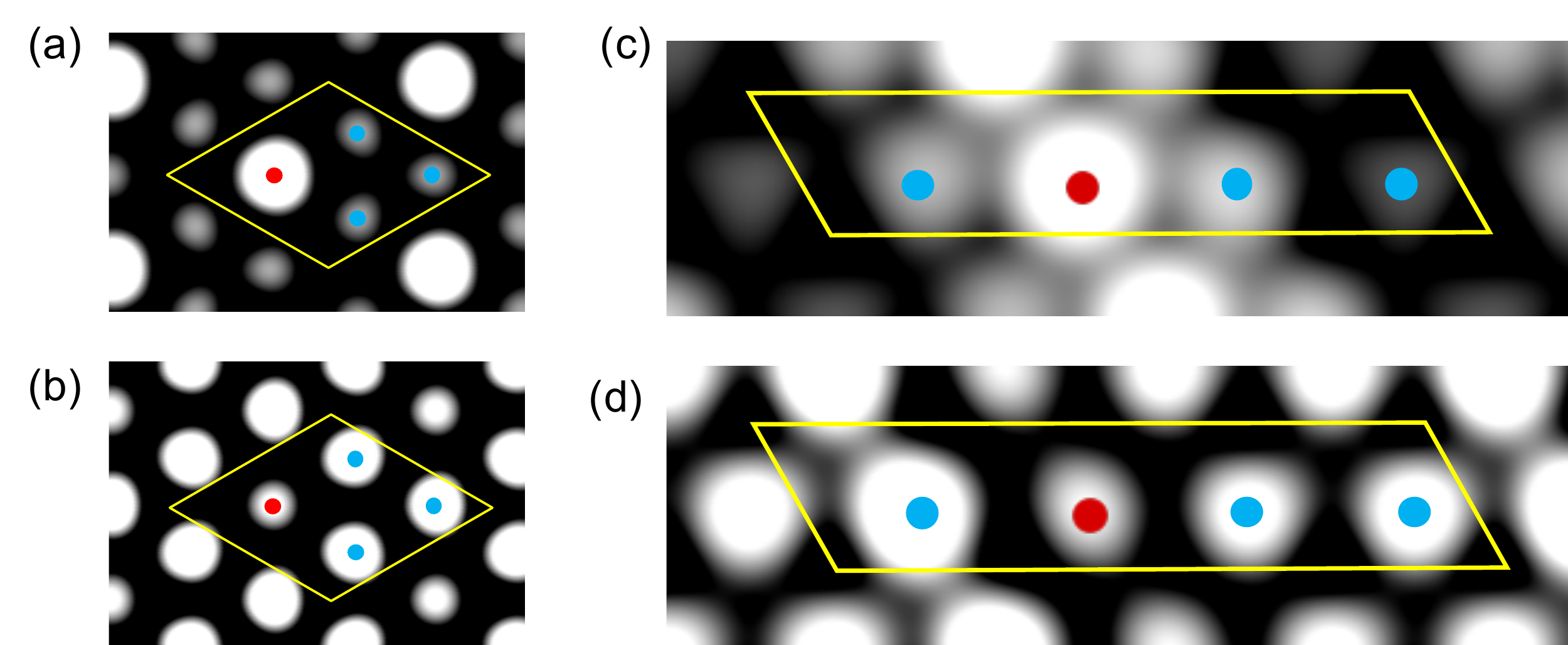}\caption{\label{fig:otherstm2}
STM simulations for the $2\times2$ and $4\times1$ Fe-II vacancy supercells in the magnetic groundstate configuration with U equal to 0 and 4 eV. (a)$2\times2$ supercell with U=0 eV; (b)$2\times2$ supercell with U=4 eV; (c)$4\times1$ supercell with U=0 eV;(d)$4\times1$ supercell with U=4 eV.}
\end{figure}

\clearpage
\section{Liechtenstein DFT+U functional}

Table \ref{tab:liecht} shows that also for the DFT+U scheme implemented by Liechtenstein {\it et al} ~\cite{liechtenstein1995prb} the interlayer FM ordering is energetically favorable in Fe$_3$GeTe$_2$ for all listed non-zero values of the Hubbard $U$ and Hund's coupling $J_H$ parameters.

\begin{table}[!htb]
\begin{centering}
\begin{tabular}{ccc}
\hline 
U (eV) & $J_{H}=0.15$ & $J_{H}=0.60$\tabularnewline
\hline 
\hline 
0 & 30.0 & 16.4\tabularnewline
\hline 
1 & -26.4 & -4.9\tabularnewline
\hline 
2 & -45.7 & -40.0\tabularnewline
\hline 
3 & -35.3 & -37.6\tabularnewline
\hline 
4 & -39.6 & -34.4\tabularnewline
\hline 
\end{tabular}
\par\end{centering}
\caption{\label{tab:liecht} The total energy difference per cell (in units of meV) between interlayer FM and AFM ordering in the normal cell without Fe-II vacancies for varying values of the Hubbard $U$ and Hund's coupling $J_H$.}
\end{table}

\clearpage
\section{Te heights}

Table \ref{tab:height} shows that for all the considered Fe-II vacancy supercells and values of the Hubbard U, in both the FM and AFM configuration, the Te heights above the Fe-II vacancies are lower than the Te heights above the Fe-II atoms.

\begin{table}[!htb]
\begin{centering}
\begin{tabular}{ccccccc}
\hline 
U (eV) & \multicolumn{2}{c}{$\sqrt{3}\times\sqrt{3}$ } & \multicolumn{2}{c}{$2\times2$} & \multicolumn{2}{c}{$4\times1$}\tabularnewline
\hline 
\hline 
 & FM & AFM & FM & AFM & FM & AFM\tabularnewline
\hline 
0 & 8.3 & 8.4 & 8.1 & 7.6 & 10.1 & 9.3\tabularnewline
\hline 
1 & 8.8 & 9.9 &  &  &  & \tabularnewline
\hline 
2 & 10.7 & 12.6 &  &  &  & \tabularnewline
\hline 
3 & 22.4 & 20.0 &  &  &  & \tabularnewline
\hline 
4 & 24.9 & 23.5 & 16.0 & 15.2 & 10.8 & 9.4\tabularnewline
\hline 
\end{tabular}
\par\end{centering}
\caption{\label{tab:height} The relative Te heights (in units of pm) between that above the Fe-II atoms(average) and Fe-II vacancies for three Fe-II vacancy supercells and values of the Hubbard U obtained from relaxing the atomic positions within the DFT+U method~\cite{dudarev1998electron}. The empty entries correspond to parameter values that have not been considered.}
\end{table}

\clearpage
\section{Integrated Te-p Density of States}

Table \ref{tab:intdos} shows that for the AFM DFT simulation without a Hubbard U correction the integrated DOS of the Te-$p$ orbitals above the Fe-II vacancies is higher than the integrated DOS of the Te-$p$ orbitals above the Fe-II atoms. For the FM DFT+U simulations with $U=4$eV this is reversed.

\begin{table}[!htb]
\begin{centering}
\begin{tabular}{ccccccc}
\hline 
U (eV) & \multicolumn{2}{c}{$\sqrt{3}\times\sqrt{3}$} & \multicolumn{2}{c}{$2\times2$} & \multicolumn{2}{c}{$4\times1$}\tabularnewline
\hline 
\hline 
 & Fe-II atom & Fe-II vacancy & Fe-II atom & Fe-II vacancy & Fe-II atom & Fe-II vacancy\tabularnewline
\hline 
0 & 0.077 & 0.144 & 0.084 & 0.119 & 0.068 & 0.126\tabularnewline
\hline 
4 & 0.160 & 0.083 & 0.128 & 0.121 & 0.139 & 0.119\tabularnewline
\hline 
\end{tabular}
\par\end{centering}
\caption{\label{tab:intdos} The (average) Density of States of the Te-$p$ orbitals above the Fe-II atoms and Fe-II vacancies integrated between the STM bias voltage and the Fermi energy, obtained from the AFM DFT simulation without a Hubbard U correction and the FM DFT+U simulations with $U=4$eV for three Fe-II vacancy supercells.}
\end{table}

\end{widetext}

\end{document}